\documentclass[twocolumn,showpacs,preprintnumbers,amsmath,amssymb]{revtex4}

\usepackage{graphicx}
\usepackage{dcolumn}
\usepackage{bm}

\begin{document}


\title{Partial Synchronization on Complex Networks}
\author{Bin Ao and Zhigang Zheng\footnote{Corresponding author: zgzheng@bnu.edu.cn.}}
\address{
Department of Physics and the Beijing-Hong-Kong-Singapore Joint
Center for Nonlinear and Complex Systems (Beijing), Beijing Normal
University, Beijing 100875, CHINA }

\begin{abstract}
Network topology plays an important role in governing the collective
dynamics. Partial synchronization (PaS) on regular networks with a few
non-local links is explored. Different PaS patterns out of the
symmetry breaking are observed for different ways of non-local
couplings. The criterion for the emergence of PaS is studied. The
emergence of PaS is related to the loss of degeneration in
Lyapunov exponent spectrum. Theoretical and numerical analysis
indicate that non-local coupling may drastically change the
dynamical feature of the network, emphasizing the important
topological dependence of collective dynamics on complex networks.
\end{abstract}

\maketitle

Synchronization, as a universal cooperative behavior and a
fundamental mechanism in nature, has been extensively studied in
relating to numerous phenomena in physics, chemistry, and biology
\cite{1,2,3,4}. In recent years, there has been a growing interest
in the synchronization of spatiotemporal systems, especially in
synchronous dynamics on networks. Synchronization on typical
complex networks, e.g., on small-world networks \cite{5},
scale-free networks \cite{ab,jalan} or weighted
network\cite{hwang} have been investigated recently\cite{6}. In
spite of these efforts, a number of fundamental questions still
remain open. For example, most of previous works have been focused
on the onset of global synchronization, while much less was
explored for partial synchronization prior to the global case.
However, there exist rich synchronous dynamics on complex
networks, and many of these behaviors are closely related to
partially synchronous motions. Moreover, the mechanism for
synchronization on complex networks is still not clear. A good
understanding of this issue should be relevant to many collective
behaviors in spatiotemporal systems, especially in complex
networks.

An important feature of complex networks is the existence of
non-local links. On the one hand, dynamics on general complex
networks (e.g., small-world or scale-free types) are usually
difficult to study due to the sophisticated network topology; On
the other hand, simple networks, especially regular networks,
cannot be ideal models in describing real networks. Therefore, one
should find some kind of network that is relatively simple but
complex enough to reflect properties of typical complex networks.
It is intuitive and feasible to consider a regular network plus
only a few non-local links, which forms the simplest "complex"
network. This can be regarded as a good bridge between regular
networks and general complex networks. In this Letter, we address
the issue of synchronization on rings of coupled oscillators with
a few shortcut links. We reveal important partial synchronization
(PaS) phenomena induced by symmetry breaking in the presence of
non-local couplings. Various synchronous patterns are observed for
different types of non-local links. This reveals the significant
role of network topology in governing the global dynamics. We
analyze in detail the conditions for the emergence of PaS and give
the criterion of the topological dependence of PaS.

We apply certain $m$-dimensional dynamics to nodes in a given network:
$\dot{\vec{x}}=\vec{f}(\vec{x}),\vec{x}=(x_1,x_2,\cdots,x_m)$. The
network is formed by using the linear coupling as edges. The
dynamical equations of the network can be written as
\begin{eqnarray}
\dot{\vec{X}}=\vec{F}(\vec{X})+\varepsilon\Gamma\otimes C \vec{X},
\label{e1}
\end{eqnarray}
where $\vec{X}=(\vec{x}^1,\vec{x}^2,\cdots,\vec{x}^N)$,
$\vec{F}=(\vec{f}^1,\vec{f}^2,\cdots,\vec{f}^N)$, and $N$ is the
number of nodes. $\varepsilon$ denotes the coupling strength, and
$\Gamma:\mathbb{R}^m \rightarrow \mathbb{R}^m$ characterizes the
coupling scheme among the nodes. $C=M-D$, where $M$ is the
adjacency matrix of the network (the element $M_{ij}$ denotes the
number of the edges that link node $i$ and $j$). $D$ is a diagonal
matrix, and $D_{ii}=\sum_{j=1}^{N}M_{ij}$. Therefore
$\sum_{j=1}^{N}C_{ij}=0$, and its largest eigenvalue $\lambda_1^C$
is 0 (we adopt $\lambda_1\geq\lambda_2\geq\cdots\geq\lambda_N$
throughout this Letter).

\begin{figure}
\includegraphics[width=3.5in,height=1.7in]{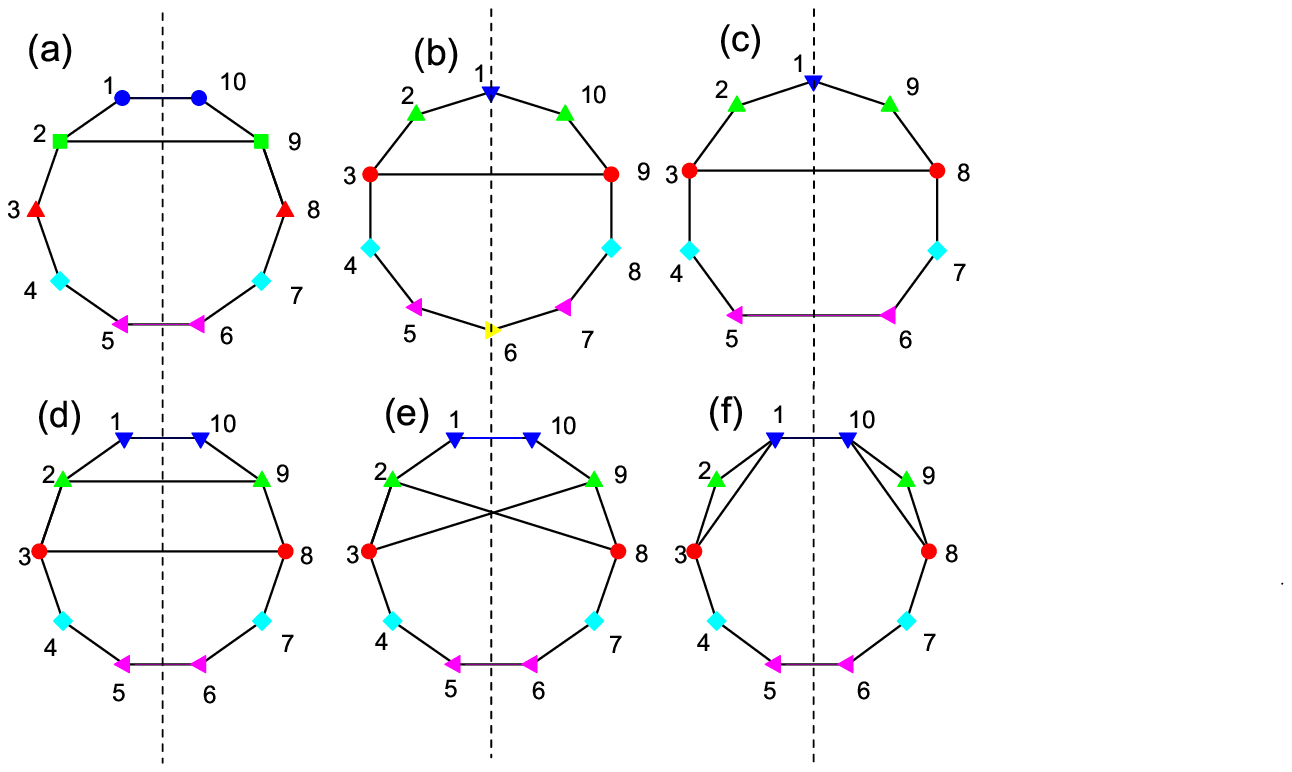}
\caption{\label{f1}Different Topologies of regular networks with
non-local links. (a): One shortcut leads to five synchronized
clusters(10 nodes); (b): One shortcut leads to six clusters (10
nodes).(c): One shortcut leads to five clusters (9 nodes). (d):
Two parallel shortcuts; (e)Two crossed shortcuts; (f)The
lambda-type network.}
\end{figure}

We first briefly describe the synchronous phenomena we observed in
the above networks. Above all, one can define the average distance
between two oscillators $i$ and $j$ by $d_{ij}=\lim_{T\to
\infty}(T^{-1}\int\|\vec{x_i}-\vec{x_j}\|dt)$, and say they are in
a synchronous state in the sense that $d_{ij}\to0$. Denote the
number of non-local links by $N_{s}$. When $N_{s}=1$, the
synchronous patterns are shown in Figs.~\ref{f1}(a)-(c), where the
nodes with the same symbol will synchronize to each other, forming
several synchronous clusters. We call the formation of these
synchronous clusters "partial synchronization" (PaS). Several
interesting phenomena can be observed: (1) PaS emerges not only
for oscillators directly linked by the non-local couplings, but
also for those without shortcuts; (2) Synchronous clusters are
formed by satisfying the mirror symmetry of lattices, i.e., if
$(i,j)$ is a pair linked by a shortcut, then $(i+k,j-k)$ are all
synchronous pairs. More interestingly, if there are two non-local
links, the emergence of PaS also need not only satisfy the mirror
symmetry of network, but also depend on the linking form. For
example, PaS can always be achieved on networks with "parallel"
links [the network with two shortcuts $(i,j)$ and $(i+k,j-k)$], as
shown in Fig.~\ref{f1}(d). However, for networks with "cross"
shortcuts [two shortcuts $(i,j)$ and $(i+k,j+k)$] shown in
Fig.~\ref{f1}(e) and networks with "lambda"-type shortcuts [two
shortcuts $(i,i+k)$ and $(j,j+k)$] shown in Fig.~\ref{f1}(f), PaS
emerges only for some choices of $k$. If PaS emerges on "cross"
and "lambda" networks, synchronization may not emerge between
pairs that are directly linked.

The above PaS behavior implies the important role of network
topology in governing the collective behaviors. It is important to
understand the mechanism of the above interesting PaS phenomena.
For example, why does PaS appear on networks with mirror symmetry
of sites? What is the manifestation in the dynamical exponents at
the onset of PaS? Moreover, can we find a simple criterion in
exactly determining whether PaS can appear in a given network? It
is our central task to answer these important questions.

We start with theoretical analysis on the stability of complete
synchronization \cite{7}. For the synchronous state
$\vec{x}^1(t)=\vec{x}^2(t)=\cdots=\vec{x}^N(t)=\vec{s}(t)$, by
linearizing Eq.~(\ref{e1}) near the synchronous manifold
$\vec{s(t)}$, i.e., $\vec{x}^i=\vec{s}+\vec{u}^i
(i=1,2,\cdots,N)$, one may get
$\dot{\vec{U}}=[\vec{I}_N\otimes\emph{D}\vec{f}+\varepsilon\Gamma\otimes
C ]\vec{U}, \label{e2}$
where $\vec{U}=(\vec{u}^1,\vec{u}^2,\cdots,\vec{u}^N)$ and
$\emph{D}\vec{f}$ is the Jacobian function of $\vec{f}$ near the
synchronous state $\vec{s}(t)$. Noticing that the first term of
the new equation is block diagonal with $\emph{m}\times \emph{m}$
blocks, one can diagonalize $C$. Thus it becomes
$\dot{\vec{v}}_k=[\emph{D}\vec{f}+\varepsilon\lambda_k\Gamma
]\vec{v}_k,(k=1,2,\cdots,N), \label{e3}$
where $\lambda_k$ are eigenvalues of $C$. Then Eq.~(\ref{e1})
becomes $\emph{N}$ independent blocks . To judge whether CS can
emerge on a network, one needs to find the regime where the
largest Lyapunov exponent of the generic equation
$\dot{\vec{v}}=[\emph{D}\vec{f}+\alpha\Gamma ]\vec{v}$ is negative
on the $\alpha$ plane ($\alpha$ is complex). If all
$\varepsilon\lambda_k (k=1,2,\cdots,N)$ are in the
negative(stable) regime (except $\lambda_1=0$), then the CS state
will be stable. Therefore, the essence of this method is to divide
the system into two subsystems which dominate the dynamics on the
synchronous manifold(the equation with $\lambda_1=0$) and that on
the transversal manifold(the equation with
{$\lambda_i$}$(i=2,3,\cdots,N)$), and then the emergence of the CS
is determined by the stability of the transversal system.

For the case of PaS, one also needs the system to be divided into
two parts, and the stability of one of them (the transverse
system) determine the emergence of the PaS. On the one hand, the
symmetry of a network can be defined by the invariance under a
permutation transformation. Taking the case in Fig.~\ref{f1}(a) as
an example, its symmetry requires that $F_N C F_N^{-1}=C,$ where
$F_{N}$ is a counteridentity matrix that satisfies
$F_{i,N-i+1}=1$, for $i=1,2,\cdots,N$ and $F_{ij}=0$ for all
$i+j\neq N/2+1$, thus the network has a mirror symmetry. It is
easy to verify that PaS is a solution of the differential
equation, then each node in a cluster should possess the same
dynamics \cite{8}. Therefore the mirror symmetry is needed for the
emergence of PaS. On the other hand, one can introduce new
variables $\vec{W}=\emph{S}\vec{X}$, where $\vec{W}=(\vec{w}^1,
\vec{w}^2, \cdots, \vec{w}^{N}),$
\begin{eqnarray}
 S =\left( \begin{array}{ccc}
\emph{I}_{N/2} & F_{N/2} \\
\emph{I}_{N/2} & -F_{N/2}
\end{array}\right).
\label{e4}
\end{eqnarray}
Here $\emph{I}_{N/2}$ is a $\emph{N}$/2-order identity matrix. For
a network with a mirror symmetry, the above transformation can
divide the system into two subsystems. Although this requires the
adjacent matrix $C$ to be persymmetric (i.e., $C$ is symmetric
about its anti-diagonal), the condition can always be satisfied by
relabelling the node for a network with a mirror symmetry.
Furthermore, if a network satisfies the mirror symmetry and its
label obeys the rule in Fig.~(\ref{f1}), the adjacent matrix $C$
can be block diagonalized by the transformation (\ref{e4}):
$$\emph{G}=\emph{SCS}^{-1}
=\left(\begin{array}{cc}
A & 0 \\
0 & B
\end{array}\right).$$
This transformation only re-distributes the eigenvalue spectrum of
the matrix $C$ into two blocks $A$ and $B$. Since
$\sum_{j=1}^{N/2}A_{ij}=0$, the largest eigenvalue of $A$ is 0.

By using the above transformation and considering that we only
care about the dynamics near the synchronous manifold, the system
can be separated into two parts, and they can be linearized near
their synchronous manifold respectively:
\begin{eqnarray}
\delta
\dot{\vec{W}}_{im}=[D\vec{F}(\vec{W}_{im})+\varepsilon\Gamma
A]\delta \vec{W}_{im}, \label{e5}
\end{eqnarray}
\begin{eqnarray}
\delta\dot{\vec{W}}_{tv}=[D\vec{F}(\vec{W}_{tv})+\varepsilon\Gamma
B] \delta\vec{W}_{tv}, \label{e6}
\end{eqnarray}
where $\vec{W}_{im}=(\vec{w}^1, \vec{w}^2, \cdots, \vec{w}^{N/2})$
and $\vec{W}_{tv}=(\vec{w}^{N/2+1}, \vec{w}^{N/2+2}, \cdots,
\vec{w}^{N}).$ For the PaS case discussed above, we have
$\vec{W}_{im}=\vec{X}_1+\vec{X}_2,\vec{W}_{tv}=\vec{X}_1-\vec{X}_2$,
where $\vec{X}_1=(\vec{x}^1,\vec{x}^2,\cdots,\vec{x}^{N/2})$, and
$\vec{X}_2=(\vec{x}^{N/2+1},\vec{x}^{N/2+2},\cdots,\vec{x}^{N})$.
Here, $\vec{W}_{im}$ denotes the dynamics on synchronous manifold
and $\vec{W}_{tv}$ denotes the dynamics on transversal manifold.

Back to Eqs.(\ref{e5}) and (\ref{e6}). If one increase the
coupling from 0, synchronization will be achieved for (3) and (4)
when $\varepsilon\lambda^{im}_2$ ($\lambda^{im}_2$ is the second
largest eigenvalue of $A$) and $\varepsilon\lambda^{tv}_1$
($\lambda^{tv}_1$ is the largest eigenvalue of $B$) respectively
fall in the stable regime of the $\alpha$ plane. The
synchronization (the stability of synchronous state) of
Eq.(\ref{e6}) suggest the emergence of PaS. If
$\lambda_2^{im}<\lambda_1^{tv}$, the synchronization of
Eq.(\ref{e5}) will be achieved at a smaller coupling than that of
Eq.(\ref{e6}), thus a global synchronization state has been
achieved when Eq.(\ref{e6}) is stable. Therefore the emergence of
PaS requires
\begin{eqnarray}
\lambda_2^{im}>\lambda_1^{tv}. \label{e7}
\end{eqnarray}

\begin{figure}
\includegraphics[width=3in,height=2.1in]{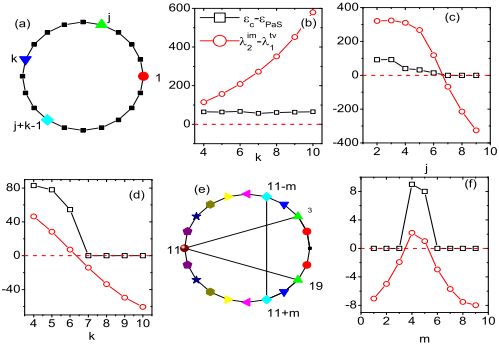}
\caption{\label{f2}(a): The label rule for $N_s=2$; (b)-(d): The
comparison between $\varepsilon_C-\varepsilon_{PaS}$ and
$\lambda_2^{im}-\lambda_1^{tv}$, N=20.
$\lambda_2^{im}-\lambda_1^{tv}$ are amplified by a proper times;
(b): Parallel network, j=3; (c): Cross network, k=9; (d): Lambda
network, j=3; (e): Topology graph of the network for $N_s=3$; (f):
The same as (b)-(d) for $N_s=3$.}
\end{figure}

Now let us test the validity of this criterion and numerically
explore the PaS phenomena. We take the Lorenz oscillator as the
node of the network: $dx/dt=\sigma(y-x)$, $dy/dt=rx-y-xz$,
$dz/dt=xy-bz$. Here $\sigma=10, r=27, b=8/3.$ We adopt the
coupling scheme by the 3$\times$3 matrix $\Gamma$ as
$\Gamma_{2,1}=1$ and $\Gamma_{i,j}=0$ for $i\neq2,j\neq1$. Let us
first try to find all possible networks that satisfy the mirror
symmetry. Taking the network with two non-local links for instance
(the case of $N_c=1$ can always topologically lead to PaS). For
convenience we always take node $i=1$ as one of the ends. The
second and the third ends are labelled as $j$ and $k$ ($k>j$),
respectively. Due to the mirror symmetry, the label of the fourth
end is $n_4=k+j-1$. There are three possible links that hold the
mirror symmetry , as shown in Figs.~\ref{f1}(d),(e) and (f).
Noticing the translational symmetry of the ring network, we have
$2\leqslant j\leqslant N/2$ and $j\leqslant k\leqslant N/2$. Thus
by varying $j$ and $k$, all symmetric networks with two non-local
couplings can be found [Fig.~\ref{f2}(a)].

\begin{figure}
\includegraphics[width=3.5in,height=1.3in]{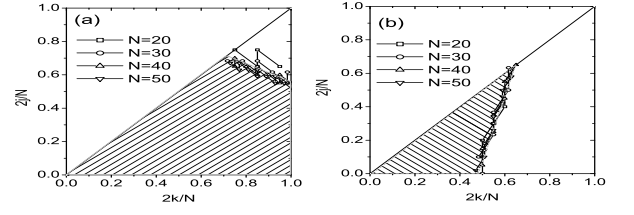}
\caption{\label{f3} The PaS regimes (labelled by sparse lines).
(a): cross network; (b): lambda network.}
\end{figure}

We use $\varepsilon_{PaS}$ and $\varepsilon_C$ to represent the
critical coupling of PaS and CS, respectively. Figs.\ref{f2}(b)
and (c) give a comparison between differences $\Delta \lambda
=\lambda_2^{im}-\lambda_1^{tv}$ and $\Delta \varepsilon
=\varepsilon_C-\varepsilon_{PaS}$ for different networks. It is
instructive to find that $\Delta \lambda$ passes zero when $\Delta
\varepsilon$ becomes zero. $\Delta \varepsilon \rightarrow 0$
means the coincidence between CS and PaS, implying that PaS
doesn't appear. This good correspondence verifies our criterion in
(\ref{e7}). Our investigations reveal that one-shortcut networks
and "parallel" networks can always achieve PaS because both
$\Delta \lambda$ and $\Delta \varepsilon$ are positive for all
$k$, as shown in Fig.\ref{f3}(b) for the "parallel" case. For
"cross" and "lambda" networks, PaS occurs only for some
configurations, as shown in Figs.\ref{f3}(c) and (d). To
illustrate these results, we plot the PaS critical line $\Delta
\lambda=0$ in the ($2j/N,2k/N$) plane. Fig.\ref{f3}(a) plots the
PaS regime for "cross" networks, and Fig.\ref{f3}(b) gives the
phase diagram of "lambda" networks. Different lines correspond to
different nodes number. These lines coincide with each other
pretty well, indicating that PaS depends on the topology of
networks and is independent of numbers of nodes.

The above criterion for PaS can also be applied to networks with
more than two non-local links, while the discussions become more
complicated. However, the above phenomena and criterion still
remain valid. Fig.\ref{f2}(f) shows one case of three shortcuts.
There are 20 nodes in a network, and two non-local links are fixed
to be (11,3) and (11,19), and the third link is (11+$m$,11-$m$),
where $m=1,2,\cdots,9$, see Fig.\ref{f2}(e). PaS occurs only for
the $m=4, 5$ cases, as shown in Fig.\ref{f2}(f).

\begin{figure}
\includegraphics[width=3in,height=2in]{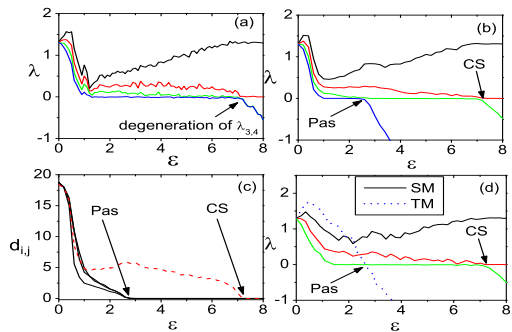}
\caption{\label{f4} (a): The first four largest LEs of a ring
network; (b): The first four largest LE's of a ring network with
one shortcut (1,4); (c): average trajectory distances
$d_{1,4},d_{2,3},d_{5,6}$ (solid lines), and
$d_{1,2},d_{3,4},d_{1,6},d_{4,5}$ (dashed lines) varying with the
coupling, where the network is the same as in (b); (d): The three
largest LE's of the synchronous manifold (solid lines) and the
largest LE of the transversal manifold.}
\end{figure}

Let's observe the manifestation of the Lyapunov exponent (LE)
spectrum (LES) of the network at the onset of PaS. Fig.\ref{f4}(a)
gives the first four largest LE's for a ring of 6 nodes without
shortcuts. One can find that the third and the fourth largest LE's
pass zero at the same coupling ($\varepsilon=\varepsilon_C$),
indicating the degeneration of LE's at the onset of CS for ring
network. Fig.\ref{f4}(b) is the four largest LE's of a ring
network with one non-local link (1,4). Obviously the presence of
non-local links leads to a breaking of degeneration of LES. The
fourth largest LE passes zero at a very small coupling. It also
can be found that $d_{1,4},d_{2,3},d_{5,6}$ become 0 at the same
coupling, as shown in Fig.\ref{f4}(c), but
$d_{1,2},d_{3,4},d_{1,6},d_{4,5}\neq 0$. Therefore, PaS emerges if
and only if there only are two positive LE's. This conclusion was
tested to be true for different node numbers and different
shortcut numbers and link ways.

These results can be interpreted by analyzing eigenvalues of the
adjacent matrix. The adjacent matrix of a ring network is a
circulant matrix, its eigenvalues are conjugated besides the
largest and the smallest ones. So when the coupling increases,
$\varepsilon\lambda_2$ and $\varepsilon\lambda_3$ will fall into
the stable regime at same critical coupling, above which the third
and fourth LE's corresponding to $\lambda_2$ and $\lambda_3$
become negative (there are always at least one positive LE and one
zero LE for a chaotic system). When non-local couplings are
applied, the degeneration in the eigenvalues and LES is lost.
Since the emergence of PaS requires
$\lambda_2^{im}>\lambda_1^{tv}$, when $\varepsilon\lambda_1^{tv}$
is in stable regime, there are only two
eigenvalues($\lambda_1^{im}=0$ and $\lambda_2^{im}$) outside of
the stable regime. That is, there are only two positive LE's. One
can also calculate the conditional LE spectrum of the synchronous
manifold (Eq.(\ref{e5})) and its corresponding transversal
manifold (Eq.(\ref{e6})). It can be found from Fig.\ref{f4}(d)
that there are only two positive conditional LE's of synchronous
state(solid line) when the largest conditional LE of the
transversal manifold (dotted line) passes zero.

In conclusion, in this letter we explored synchronization
behaviors of complex networks with a few non-local links. The
criterion for the emergence of PaS is theoretically given and
verified in numerical simulations. The manifestation of Lyapunov
exponent spectrum is discussed for PaS. Our results and criterion
proposed in this Letter are general, and they are independent of
the dynamics on sites and coupling forms. we tested different
dynamics on the nodes, such as Lorenz oscillators, Rossler
systems, and nonlinear maps (e.g., the logistic map and circle
maps). All these numerical experiments support our criterions and
conclusions. These "simple" complex networks are not only simple
but also typical, and it can be a useful tool to study more
complex networks. The synchronization behavior we studied in this
letter reveal the importance of network topology. Different types
of shortcuts may drastically change the topology and furthermore
the dynamics of a network. Our work should illuminate a good step
in understanding the synchronization processes on general complex
networks from a dynamical point of view.

\begin{acknowledgments}
We thank Prof. Gang Hu for constructive comments and suggestions,
and we are also grateful to Drs. H.Chen, Z.Cao, N.Zheng and X.Feng
for very useful discussions and help.
\end{acknowledgments}


\begin{thebibliography}{99}
\bibitem{1}L. M. Pecora and T. L. Carroll, Phys. Rev. Lett. 64,
821(1990); A. Pikovsky, M. Rosenblum and J. Kurths,
\textit{Synchronization: A universal concept in nonlinear
sciences} (Cambridge University Press, Cambridge, 2001); S.
Boccaletti, C. Grebogi, Y. C. Lai, H. Mancini, and D.Maza, Phys.
Rep. 329, 103 (2000). S. Boccaletti, J. Kurths, G. Osipov, D. L.
Valladares, C. S. Zhou, Phys. Rep. 366, 1 (2002).
\bibitem{2}M. G. Rosenblum, A. S. Pikovsky, and J. Kurths Phys. Rev. Lett. 76, 1804
(1996).
\bibitem{3}Z. Zheng, G. Hu, and B. Hu Phys. Rev. Lett. 81, 5318
(1998). U. Ernst \textit{et al.}, \textit{ibid} 74, 1570 (1995).
G. Hu \textit{et al.}, \textit{ibid} 85, 3377 (2000).
\bibitem{4}N. Parekh, S. Parthasarathy, and S. Sinha Phys. Rev. Lett. 81, 1401
(1998).
\bibitem{5}D. J. Watts and S. H. Strogatz, Nature (London) 393, 440 (1998).
\bibitem{ab}S.H. Strogatz, Nature 410, 268 (2001); R. Albert and A.L. Barabasi, Rev. Mod. Phys. 74, 47 (2002);
M.E.J. Newman, SIAM Rev. 45, 167(2003).
\bibitem{jalan}S. Jalan and R. E. Amritkar Phys. Rev. Lett. 90 014101 (2003).
\bibitem{hwang}D.-U. Hwang, M. Chavez \textit{et al.},
Phys. Rev. Lett. 94, 138701 (2005). M. Chavez, D.-U. Hwang
\textit{et al.}, Phys. Rev. Lett. 94, 218701 (2005).
\bibitem{6}M. E. J. Newman \textit{et al.}, Phys. Rev. Lett. 84, 3201
(2000); L. F. Lago-Fernandez \textit{et al.}, \textit{ibid} 84,
2758 (2000); M. Barahona and L. M. Pecora, \textit{ibid} 89,
054101 (2002); G. W. Wei \textit{et al.}, \textit{ibid} 89,
284103(2002); T. Nishikawa \textit{et al.}, \textit{ibid} 91,
014101(2003); F. M. Atay \textit{et al.}, \textit{ibid} 92,
144101(2004); Y. Jiang \textit{et al.}, Phys. Rev. E 68,
065201(R)(2003); J. G. Restrepo \textit{et al.}, Phys. Rev. E 69,
066215(2004); Y. Moreno and A. F. Pacheco, Europhys. Lett. 68,
603(2004).
\bibitem{siam}I. Stewart, M Golubitsky and M. Pivato, SIAM J.
Applied Dynamical Systems 2(4) 609 (2003)
\bibitem{7}J. Yang \textit{et al.}, Phys. Rev. Lett. 80, 496
(1998); L. M. Pecora and T. L. Carroll, \textit{ibid} 80, 2109
(1998).
\bibitem{8}Y. Zhang \textit{et al.}, Phys. Rev. E 63,026211(2001).
\end{thebibliography}
\end{document}